\documentclass[journal]{IEEEtran}
\usepackage{graphicx}
\usepackage{booktabs}
\usepackage{subfigure}
\usepackage{fontawesome}
\usepackage{overpic}
\usepackage{amsmath}
\usepackage{amssymb}
\usepackage{multirow}
\usepackage{makecell}
\usepackage{array}
\usepackage{stfloats}
\usepackage{listings}
\usepackage{algorithm}
\usepackage{color, soul}
\usepackage{bbding}
\usepackage{xcolor}
\usepackage{graphicx}
\usepackage{algpseudocode}
\usepackage{tabularx}
\usepackage{hyperref}
\hypersetup{
    colorlinks=true,
    linkcolor=blue,
    urlcolor=blue,
    citecolor=blue
}

\begin{document}
\title{Adaptive Frequency Enhancement Network for Remote Sensing Image Semantic Segmentation}
\author{
  Feng Gao, \emph{Member}, \emph{IEEE},
  Miao Fu, \emph{Student Member}, \emph{IEEE},
  Jingchao Cao,
  Junyu Dong, \emph{Member}, \emph{IEEE},\\
  Qian Du, \emph{Fellow}, \emph{IEEE}
\thanks{This work was supported in part by the National Science and Technology Major Project under Grant 2022ZD0117202, in part by the Natural Science Foundation of Shandong Province under Grant ZR2024MF020, and in part by the Natural Science Foundation of Qingdao under Grant 23-2-1-222-ZYYD-JCH. (Corresponding author: Jingchao Cao.)

Feng Gao, Miao Fu, Jingchao Cao, and Junyu Dong are with the School of Computer Science and Technology, Ocean University of China, Qingdao 266100, China. 

Qian Du is with the Department of Electrical and Computer Engineering, Mississippi State University, Starkville, MS 39762 USA.}}

\markboth{IEEE Transactions on Geoscience and Remote Sensing}
{Shell}

\maketitle

\begin{abstract}

Semantic segmentation of high-resolution remote sensing images plays a crucial role in land-use monitoring and urban planning. Recent remarkable progress in deep learning-based methods makes it possible to generate satisfactory segmentation results. However, existing methods still face challenges in adapting network parameters to various land cover distributions and enhancing the interaction between spatial and frequency domain features. To address these challenges, we propose the Adaptive Frequency Enhancement Network (AFENet), which integrates two key components: the Adaptive Frequency and Spatial feature Interaction Module (AFSIM) and the Selective feature Fusion Module (SFM). AFSIM dynamically separates and modulates high- and low-frequency features according to the content of the input image. It adaptively generates two masks to separate high- and low-frequency components, therefore providing optimal details and contextual supplementary information for ground object feature representation. SFM selectively fuses global context and local detailed features to enhance the network's representation capability. Hence, the interactions between frequency and spatial features are further enhanced. Extensive experiments on three publicly available datasets demonstrate that the proposed AFENet outperforms state-of-the-art methods. In addition, we also validate the effectiveness of AFSIM and SFM in managing diverse land cover types and complex scenarios. Our codes are available at \url{https://github.com/oucailab/AFENet}.

\end{abstract}

\begin{IEEEkeywords}
Semantic segmentation, 
remote sensing, 
frequency domain features, 
adaptive feature interaction,
spatial-frequency integration.
\end{IEEEkeywords}

\IEEEpeerreviewmaketitle

\section{Introduction}

\IEEEPARstart{W}{ITH} the rapid advancement of satellite and aerial sensor technology, an increasing number of high-resolution remote sensing images are now being acquired \cite{drm22tgrs} \cite{lz24tgrs} \cite{zhuh24tgrs}. These images play a crucial role in a wide range of applications, including land use monitoring \cite{pj24jstars} \cite{qkl15grsl}, coastal zone management \cite{cjy21jstars}, crop health assessment \cite{ly22jstars} \cite{zhu24tgrs4}, and semantic segmentation \cite{psm21grsl}. Among these applications, semantic segmentation of high-spatial resolution remote sensing images is particularly essential for interpreting and comprehensively understanding land cover \cite{myf24grsl} \cite{segnet17tpami} \cite{clc18tpami}. Therefore, this paper primarily focuses on designing an efficient, deep learning-based method for the semantic segmentation of high-spatial-resolution remote sensing images.

In the past decades, many segmentation methods based on convolutional neural networks (CNNs) have been extensively employed for remote sensing image interpretation. Dilated convolution \cite{clc18tpami}, fully convolutional network (FCN) \cite{long15fcn}, pyramid feature modeling \cite{zpp20tip}, semantic decoupling \cite{zcy23tgrs}, and memory preserving network \cite{xjj24tgrs} have been applied for remote sensing image semantic segmentation.

\begin{figure}[]
\centering  
\includegraphics[width=0.4\textwidth]{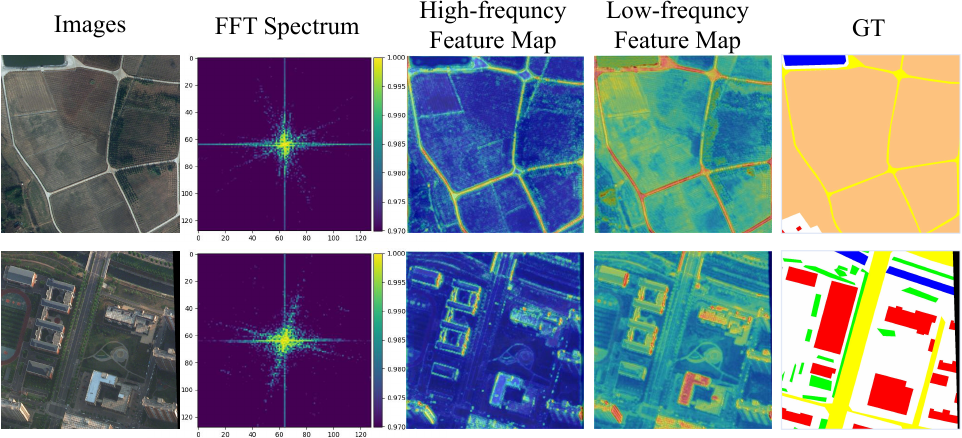}   
\caption{Results of frequency separating in remote sensing images from the LoveDA dataset across rural (first row) and urban (second row) scenes. The first column shows the original images, and the second column presents the corresponding FFT spectra, reflecting the frequency characteristics of different scenes caused by variations in object types and texture distributions. The third and fourth columns display high-frequency and low-frequency feature maps, respectively, obtained via our dynamic window mechanism.}
\label{fig1}
\end{figure}

Through multiple layers of convolution, CNN-based methods can progressively capture spatial hierarchies that are crucial for distinguishing different land cover types \cite{zh24tgrs3}. However, the limited receptive field of CNN-based methods constrains their capability for global context modeling. In recent years, the Transformer-based methods have attracted widespread attention from the remote sensing community \cite{xjw24grsl} \cite{lxt25tmm}. Self-attention \cite{sr21iccv}, pyramid Transformer \cite{zwq22cvpr}, large kernel attention \cite{gmh22neurips}, and Swin Transformer \cite{hx22tgrs} \cite{wl22grsl} \cite{zx24jstars} have been employed for remote sensing image segmentation.

Advanced deep learning-based semantic segmentation models have demonstrated remarkable success. However, it is non-trivial to design robust segmentation methods for various ground objects due to the following two challenges: 1) \textbf{Limitations in adapting network parameters to the input data with different land covers.} For different input remote sensing images, there are significant variations in the types and distributions of land cover. As illustrated in Fig. \ref{fig1}, rural area (first row) consists of smooth-textured vegetation, while urban area (second row) is covered by buildings with rich texture details. Most existing methods use the same network parameters for all input images, which limits the representation power. 2) \textbf{The interactions between frequency and spatial features need to be enhanced.} Most existing CNN-based methods do not delve into the distinct frequency variations inherent to different land cover types. Enhancing the interactions between spatial and frequency features may improve the feature representation power. Hence, how to modulate spatial and frequency features poses the second challenge.

To address the aforementioned challenges, we propose \textbf{A}daptive \textbf{F}requency \textbf{E}nhancement \textbf{Net}work, dubbed as AFENet, for high-resolution image semantic segmentation. Specifically, AFENet consists of two key components: the Adaptive Frequency and Spatial feature Interaction Module (AFSIM) and the Selective feature Fusion Module (SFM). AFSIM adaptively separates and modulates frequency-domain features to dynamically emphasize informative high- and low-frequency components corresponding to different land cover types. As shown in Fig. \ref{fig1}, the proposed AFFNet adaptively enhances the high-frequency component for urban areas, while suppressing the high-frequency component for rural areas. In addition, SFM integrates context and detail features by selectively fusing multi-scale information, enhancing the overall representation capability of the network. To demonstrate the effectiveness of our proposed method, we conduct extensive experiments on three publicly available datasets. The experimental results show the proposed AFENet outperforms the state-of-the-art methods and has strong scalability to different land cover types.


In summary, the contributions of this work are three-fold:

\begin{enumerate}

\item We proposed a novel remote sensing image segmentation framework, AFENet, which leverages both frequency and spatial domain information to effectively model ground object features. Frequency and spatial domain features are integrated and modulated via cross-attention and selective feature fusion. 

\item We designed the Adaptive Frequency and Spatial feature Interaction Module (AFSIM), which adaptively modulates the high- and low-frequency components according to the content of the input image. It adaptively generates two masks to separate high- and low-frequency components for each input image, therefore providing optimal details and contextual supplementary information for ground object feature representation. 

\item We conducted extensive comparative experiments on three publicly available datasets to validate the effectiveness of AFENet. As a side contribution, we will release our code to benefit other researchers in remote sensing image semantic segmentation community.
\end{enumerate}

\section{Related Works}

\subsection{CNN-Based Remote Sensing Image Semantic Segmentation}

In the 2012 ImageNet Large Scale Visual Recognition Challenge (ILSVRC), Krizhevsky et al. \cite{krizhevsky2012imagenet} demonstrated the potential of CNNs for image understanding. In 2016, the Fully Convolutional Network (FCN) \cite{long15fcn} replaced fully connected layers with convolutional layers, achieving end-to-end remote sensing image semantic segmentation \cite{an16accv}. However, the decoder of FCN can hardly recover the image details, leading to less precise segmentation results. To solve this problem, U-Net \cite{ronneberger2015u} has been very popular for remote sensing image segmentation. It uses the skip connections to preserve spatial information lost during downsampling. However, the limited receptive field of these methods constrains their capability for global context modeling. 

To enlarge the receptive field of CNN-based methods, Li et al. \cite{LI202184} proposed an attentive bilateral contextual network for remote sensing image semantic segmentation, in which a contextual path is built to capture the global contextual information. Yu et al. \cite{yb18jstars} employed the pyramid pooling module to extract multi-scale features. Zhao et al. \cite{zq22tgrs} proposed a pyramid attention module for multi-scale feature refinement. Although these methods alleviated the limited receptive field problem to some extent, these methods can hardly capture global contextual information. 

\subsection{Tansformer-Based Remote Sensing Image Semantic Segmentation}

Transformer-based methods have shown promising capabilities in remote sensing image interpretation \cite{ssmae24tgrs}. They use the self-attention mechanisms that allow them to capture long-range dependencies in images. It is particularly beneficial for remote sensing, where ground objects vary significantly in size and shape. Xie et al. \cite{xie2021segformer} introduced the SegFormer framework, which generates hierarchical feature maps by merging overlapping patches. The proposed multi-layer perceptron decoder within the framework adaptively integrates information from multiple layers. Wang et al. \cite{wang2022novel} employed a block-based segmentation approach, dividing the input image into non-overlapping blocks and processing them through cascaded Swin Transformer \cite{liu2021swin} to capture multi-scale features. To address the scarcity of local features when using Transformers as encoders, Xu et al. \cite{xu2023rssformer} embedded the Transformer with CNN, adaptively reducing background noise through multi-scale features. Zhang et al. \cite{zj24tgrs} proposed a dual-domain Transformer for remote sensing image segmentation. Ma et al. \cite{mxp24tgrs} proposed a CNN and Transformer hybrid network for remote sensing semantic segmentation. It adaptively employed mutually boosted attention layers and self-attention layers alternately to learn cross-modality representations of high interclass separability and low intraclass variations. Zhang et al. \cite{zyj24tgrs} presented a global adaptive second-order Transformer network for remote sensing segmentation. Feature shift module and hierarchical feature fusion module are used for global feature enhancement. Ai et al. \cite{ajq24tgrs} proposed a pyramid vision Transformer assisted by the long-time AIS data for sea-land segmentation. Liu et al. \cite{lws24tgrs} proposed a feature interaction module to combine local features with global representations extracted by Transformers and CNN, respectively.

\begin{figure*}[t]
\centering
\includegraphics [width=6in]{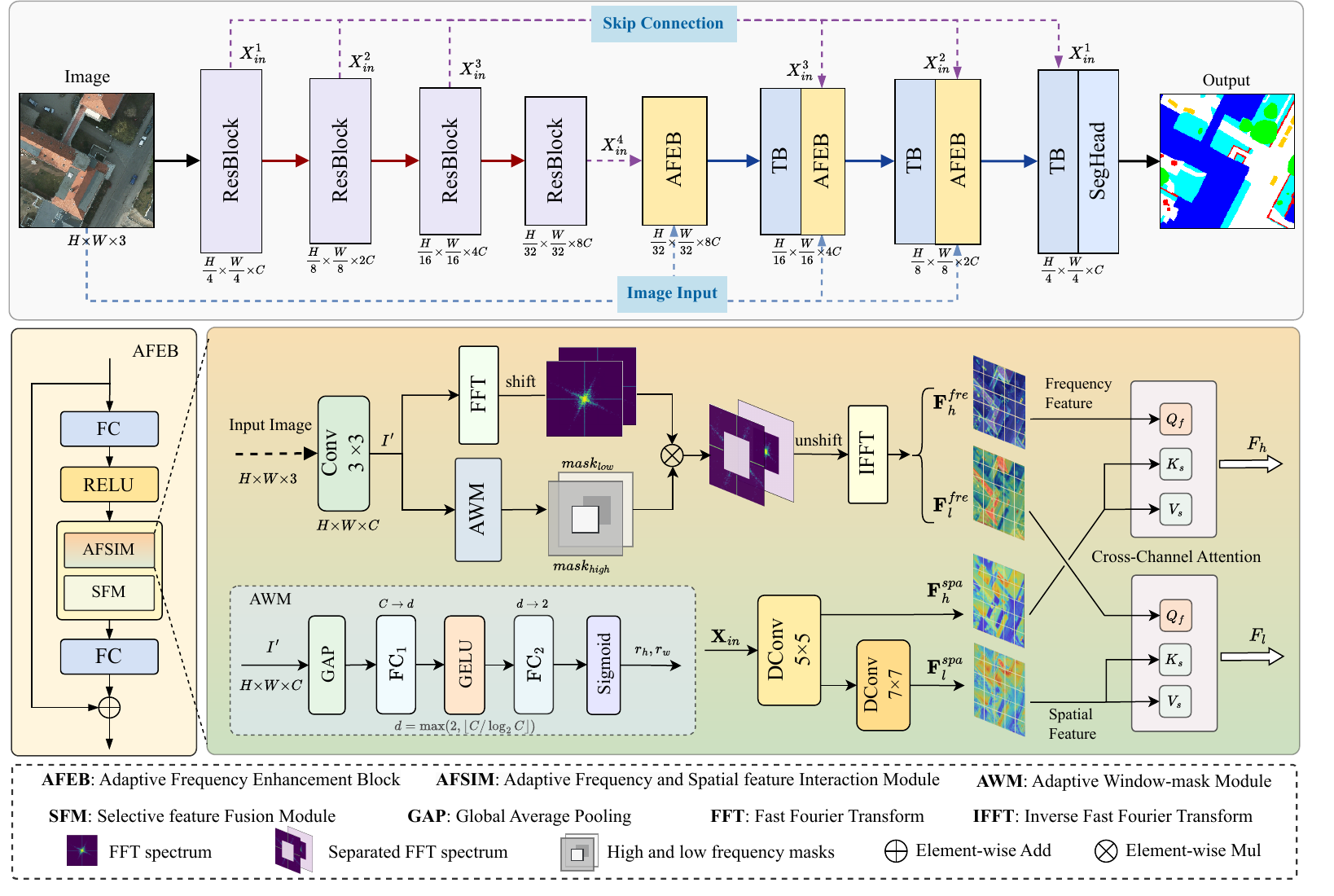}
\caption{Overview of the proposed AFENet framework. The Adaptive Frequency Enhancement Block (AFEB) module acquires high-frequency and low-frequency features through an adaptive frequency separation mechanism and employs a cross-attention mechanism to facilitate the interaction between spatial and frequency domain features. The Transformer Block (TB) further refines and completes contextual information, while promoting deeper feature integration.}
\label{fig_framework}
\end{figure*}
\subsection{Frequency Domain Feature Learning}

Recently, the frequency-domain features have been employed in deep neural networks for feature learning \cite{ych24grsl} \cite{lzy24grsl} \cite{wy22grsl} \cite{qxf22grsl} \cite{zhu24classification}. The seamless integration of frequency-domain features in deep networks has facilitated the learning of non-local feature representation \cite{li21fno}. The Fourier transform offers global frequency analysis by decomposing an image into a sum of sinusoidal and cosine functions at different frequencies. Shan et al. \cite{shan2021decouple} used the Fourier transform to decompose images into high-frequency and low-frequency components for semantic segmentation. Yang et al. \cite{yang2020fda} applied FFT to extract low-frequency spectra in a self-supervised segmentation task, reducing the gap between synthetic and real datasets. Bai et al. \cite{bai2022msanlfnet} designed a non-local filtering module to capture global contextual information in the Fourier domain, improving segmentation accuracy by integrating spatial and frequency information. Zhang et al. \cite{zhang2023fsanet} employed a frequency-based self-attention mechanism using 2D DCT for low-frequency information. Recent frequency-domain methods have significantly improved remote sensing image segmentation. For instance, Zhang et al. \cite{10601227} combined spatial convolution and attention with frequency-domain FFT in a dual-domain transformer. Liu et al. \cite{10707598} integrated deformable convolution with Fourier transforms for road segmentation. Li et al. \cite{10847802} proposed a frequency decoupling network that separately optimizes high-frequency and low-frequency features.

To effectively utilize frequency domain information for the semantic segmentation of remote sensing images, we designed an adaptive high-frequency and low-frequency information separation method based on the FFT. This method aims to enhance deep learning models' understanding of both large-scale semantic information and fine details, such as edges, in remote sensing images. Unlike conventional frequency information separation methods, our module adaptively calculates the frequency division threshold for each remote sensing image based on the distribution of geospatial information. This allows the extraction of high-frequency and low-frequency information at different scales from the original remote sensing images. These features are then embedded into the neural network, providing interactive detail and context information to spatial features. Simultaneously, it effectively establishes the internal relationship between spatial domain features and frequency domain features.

\section{Methodology} 

This section presents the overall structure of AFENet, which utilizes a ResNet18 backbone as the encoder to extract multi-scale features. The decoder integrates Adaptive Frequency Enhancement Blocks (AFEB) to extract frequency-domain features and interact with spatial features. By adaptively separating high- and low-frequency components based on land cover distribution, AFEB enhances texture details and global context, improving segmentation accuracy. Additionally, Transformer Blocks (TB) are employed to fuse multi-scale features, ensuring both global context and local details are effectively represented in the final segmentation results.

\subsection{Proposed AFENet Framework}

The overall network structure of AFENet is illustrated in Fig. \ref{fig_framework}, comprising two parts: the encoder and the decoder. For the encoder, we employ a pre-trained ResNet18 as the backbone to extract multi-scale semantic features. ResNet18 is known for its computational efficiency and has demonstrated effectiveness in numerous remote sensing image semantic segmentation tasks \cite{LI202184} \cite{diakogiannis2020resunet} \cite{wang2022unetformer} \cite{li2021multistage}. ResNet18 consists of four stages of ResBlocks, each downsampling the feature map by a factor of 2. The features are then progressively converted into low-resolution feature maps. The channel dimensions of four feature maps are 64, 128, 256, and 512, respectively.

For the decoder, we utilize Adaptive Frequency Enhancement Block (AFEB) and Transformer Block (TB) \cite{zamir2022restormer} to progressively reconstruct the segmentation output. Between every two levels of the decoder, AFEB is inserted to adaptively refine and enhance context and detail features in both frequency and spatial domains by processing the input raw image and encoder inputs. Finally, the decoder output features are fed into a segmentation head to generate the final segmentation results. 

Since different land covers affect image content at different frequency components, we specifically designed the AFEB that extracts low- and high-frequency components from the input and then modulates them to emphasize the corresponding informative features. Next, we detail the two key components of AFEB: (1) Adaptive Frequency and Spatial feature Interaction Module (AFSIM), and (2) Selective feature Fusion Module (SFM).

\subsection{Adaptive Frequency and Spatial feature Interaction Module (AFSIM)}

To effectively mine frequency domain features in remote sensing images and supplement the context and detail features extracted by deep learning models, we designed the adaptive frequency-spatial feature interaction module (AFSIM), as illustrated in Fig. \ref{fig_framework}. Primarily, AFSIM consists of three parts: adaptive frequency separation, spatial feature fusion, and cross-domain feature interaction.

\textbf{Adaptive Frequency Separation.} To effectively extract frequency information from remote sensing images, we use domain transformation and mask generation to separate frequency information. For remote sensing images, we first downsample them to match the spatial resolution of the current layer's features, obtaining $\mathbf{X} \in \mathbb{R}^{H \times W \times 3}$. Next, a $3\times3$ convolution is applied to align $\mathbf{X}$ to the target channel dimension $C$, ensuring compatibility for subsequent frequency-domain processing. Finally, FFT is performed to transform the feature into its spectrum representation, $\mathbf{F} \in \mathbb{R}^{H \times W \times C}$.

As illustrated in Fig. \ref{fig_framework}, a 2D mask is generated by an Adaptive Window-mask Module (AWM). It computes a frequency boundary to separate the high-frequency and low-frequency information within $\mathbf{F}$. The frequency boundary is calculated based on the characteristics of the land cover distribution within each image. 

In the AWM, a neural network is used to establish an interpretable mapping from spatial features to frequency thresholds. For the input aligned channel-dimensional feature $\mathbf{X}'$, global average pooling  is first applied to obtain a global feature representation of the spatial domain, preserving the spatial and semantic information of each channel in the original feature map. Then, two linear layers ($\text{FC}_1$ and $\text{FC}_2$) with GELU activation are used to produce two ratios. $\text{FC}_1$ reduces the dimension from $C$ to $d$, and $\text{FC}_2$ maps $d$ to 2. The dimension $d$ is determined by the following formula:

\begin{equation}
d = \max\left(2, \left\lfloor C/\log_2 C \right\rfloor \right)
\label{eq:dim_compress}
\end{equation}

This design ensures smoother gradient flow and optimal information preservation. Both ratios are activated by a sigmoid function, ensuring the output values fall between 0 and 1. Overall, the process of generating thresholds in AWM can be expressed as:

\begin{equation}
[r_h, r_w] = \delta \big(\textrm{FC}_{2}\big( \sigma \big( \text{FC}_{1}(\textrm{GAP}(\mathbf{X}')) \big) \big) \big)
\end{equation}

where $\sigma$ represents the GELU activation function, $\delta$ represents the sigmoid function, and GAP denotes global average pooling. $r_h$ and $r_w$ control the height and width of the mask window, thereby adaptively defining the frequency spectrum boundaries.

\begin{algorithm}[t]
\caption{FFT-Based Adaptive Frequency Separation}
\label{alg_fft_afs}

\textbf{Input:} Feature tensor $\mathbf{X} \in \mathbb{R}^{H\times W\times 3}$

\textbf{Output:} High-frequency feature $\mathbf{F}^{fre}_h \in \mathbb{R}^{H\times W\times C}$, low-frequency feature $\mathbf{F}^{fre}_l \in \mathbb{R}^{H\times W\times C}$
\begin{algorithmic}[1]
\State $\mathbf{Initialization:} ~~ \mathbf{M}_{\text{low}} \leftarrow \mathbf{0}^{H \times W}$ 
\Statex {\textcolor{gray}{/* Initialize low-frequency mask */}}
\vspace{0.5em}
\State $\mathbf{X}' \leftarrow \textrm{Conv}_{3\times3}(\mathbf{X})$
\Statex \textcolor{gray}{ /* Align $\mathbf{X}$ to channel dimension $C$ for consistency with spatial-domain features */}
\vspace{0.5em}
\State $[r_h, r_w] \leftarrow \delta(\text{FC}_2( \sigma( \text{FC}_1(\text{GAP}(\mathbf{X}')))))$
\Statex \textcolor{gray}{/* 
    Compute adaptive ratios $r_h$/$r_w$ with dimension compression: 
    $\text{FC}_1: C \to d$ (via Eq.1), $\text{FC}_2: d \to 2$ */}
\vspace{0.5em}
\State $h \leftarrow \lfloor \frac{H}{2} \cdot r_h \rfloor,\ w \leftarrow \lfloor \frac{W}{2} \cdot r_w \rfloor$
\Statex \textcolor{gray}{ /* Compute the size of the mask window */}
\vspace{0.5em}
\State $\mathbf{M}_{\text{low}}[:, \frac{H}{2}-h:\frac{H}{2}+h, \frac{W}{2}-w:\frac{W}{2}+w] \leftarrow 1$
\Statex \textcolor{gray}{/* Generate low-frequency mask */}
\vspace{0.5em}
\State $\mathbf{M}_{\text{high}} \leftarrow \mathbf{1} - \mathbf{M}_{\text{low}}$
\Statex {\textcolor{gray}{/* Generate high-frequency mask */}}
\vspace{0.5em}
\State $\mathbf{F} \leftarrow \text{FFT}(\mathbf{X}')$
\Statex \textcolor{gray}{ /* Compute features in the frequency domain */}
\vspace{0.5em}
\State $\mathbf{F}^{fre}_h \leftarrow \textrm{InverseFFT}(\mathbf{F} \odot (1 - mask))$
\vspace{0.5em}
\State $\mathbf{F}^{fre}_l \leftarrow \textrm{InverseFFT}(\mathbf{F} \odot mask)$
\Statex \textcolor{gray}{High- and low-frequency feature separation}
\end{algorithmic}
\vspace{0.5em}
\textbf{Return} $\mathbf{F}^{fre}_h, \mathbf{F}^{fre}_l$
\end{algorithm}

Next, a window region is set based on the computed adaptive ratios $r_h$ and $r_w$. The learned thresholds physically define frequency separation boundaries as:

\begin{equation}
\mathbf{M}_{\text{low}} = \mathcal{I}\left(\frac{H}{2}-h, \frac{H}{2}+h\right) \otimes \mathcal{I}\left(\frac{W}{2}-w, \frac{W}{2}+w\right)
\label{eq:mask_def}
\end{equation}

where $\mathcal{I}(a,b)$ denotes the indicator function for coordinates $[a,b]$, $\otimes$ is the outer product, $h=\lfloor \frac{H}{2}\cdot r_h \rfloor$, and $w=\lfloor \frac{W}{2}\cdot r_w \rfloor$. The resulting window region $\mathbf{M}_{\text{low}}$ serves as the low-frequency mask.

Following that, the high-frequency mask is generated by subtracting the low-frequency mask from an all-one matrix. Both masks are applied via element-wise multiplication to the spectrum feature, yielding decoupled high-frequency and low-frequency spectrum feature maps. Finally, the inverse Fourier transform is employed to generate the low-frequency and high-frequency spatial domain features $\mathbf{F}^{fre}_l$ and $\mathbf{F}^{fre}_h$.  The corresponding pseudocode is illustrated in Algorithm \ref{alg_fft_afs}.

This adaptive frequency separation enables the model to dynamically adjust the mask size according to the content of the input image, thereby providing details and contextual supplementary information for spatial domain features.

\textbf{Spatial Feature Fusion.} The input feature $\mathbf{X}_{in}$ integrates complementary information from both the encoder and decoder. Specifically, the encoder outputs ($\mathbf{X}_{enc}$) provide multi-scale spatial feature information, while the decoder outputs ($\mathbf{X}_{dec}$) contribute to the supplementation of frequency information and contextual information.
\begin{figure}[!]
\centering
\includegraphics[width=3.5in]{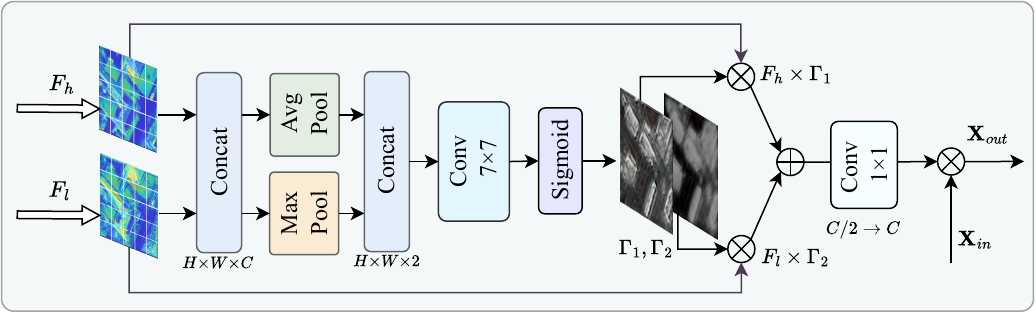}
\caption{Structure of the selective feature fusion module (SFM). High-frequency $F_h$ and low-frequency $F_l$ features are processed through average pooling and max pooling, concatenated, and used to generate a weight mask for feature fusion.}
\label{fig3}
\vspace{-3mm}
\end{figure}

To ensure consistent spatial resolution and dimensionality, we upsample the decoder features to match the resolution of the encoder features. These features are then concatenated along the channel dimension, followed by a $1 \times 1$ convolution to align the channel dimensions and reduce computational complexity. The process is defined as follows:

\begin{equation}
\mathbf{X}_{in}^i = \text{Conv}_{1 \times 1} \left( \text{Concat} \left( \text{Upsample} \left( \mathbf{X}_{dec}^{i+1} \right), \mathbf{X}_{enc}^i \right) \right)
\end{equation}
where $\mathbf{X}_{dec}^{i+1}$ represents the decoder output at layer $i+1$, and $\mathbf{X}_{enc}^i$ denotes the encoder output at layer $i$. For $i=4$, the input feature $\mathbf{X}_{in}^4$ is directly taken from the encoder's output $\mathbf{X}_{enc}^4$. The fusion mechanism unifies semantic understanding and structural details in $\mathbf{X}_{in}$, enabling comprehensive feature representation for the subsequent stages.

\textbf{Cross-Domain Feature Interaction.} After fusing complementary spatial features in $\mathbf{X}_{in}$, we apply two depth-wise convolutions with different kernel sizes to capture both contextual and detailed features of remote sensing images, each using varying receptive fields. First, a $5 \times 5$ depth-wise convolution (DWConv) is used to extract detailed features, resulting in the high-frequency spatial feature $\mathbf{F}_h^{spa}$, which preserves rich texture and edge information.

Next, a $7 \times 7$ depth-wise convolution with a larger receptive field is applied to $\mathbf{F}_h^{spa}$ to extract semantic information, yielding the low-frequency spatial feature $\mathbf{F}_l^{spa}$. The process is defined as follows:

\begin{equation}
\mathbf{F}_h^{spa} = \text{DWConv}_{5 \times 5}(\mathbf{X}_{in}),
\end{equation}
\begin{equation}
\mathbf{F}_l^{spa} = \text{DWConv}_{7 \times 7}(\mathbf{F}_h^{spa}).
\end{equation}

Frequency and spatial domain features capture different aspects of the input images. Semantic alignment is necessary to ensure consistent and complementary image representations due to their semantic differences. We adopt a multi-head cross-attention mechanism to extract shared features from both the frequency and spatial domains.

Specifically, we combine high-frequency features $\mathbf{F}^{fre}_h$ and $\mathbf{F}^{spa}_h$ together to focus on capturing local texture and edge information. Meanwhile, low-frequency features $\mathbf{F}^{fre}_l$ and $\mathbf{F}^{spa}_l$ are grouped to capture global contextual and structural information of ground objects.

To be specific, $\mathbf{F}^{fre}_h$ and $\mathbf{F}^{spa}_h$ are processed through $1\times1$ convolution followed by $3\times3$ depth-wise convolution, resulting in $Q_{f}$, $K_{s}$, and $V_{s}$. The computation is as follows:

\begin{equation}
Q_f = \text{DWConv}_{3\times3}(\text{Conv}_{1 \times 1}(\mathbf{F}^{fre}_{h})), \\
\end{equation}
\begin{equation}
K_s, V_s = \text{DWConv}_{3\times3}(\text{Conv}_{1 \times 1}(\mathbf{F}^{spa}_{h})).
\end{equation}

We then compute cross-attention as follows:
\begin{equation}
F_h = V_s \cdot \textrm{softmax}\left( \frac{K^T_s Q_f}{\sqrt{D}}\right)
\end{equation}
where $F_h$ is the features enhanced by the high-frequency domain information. $D$ is the number of hidden dimensions for a single head in the attention.

Similarly,  $\mathbf{F}^{fre}_l$ and $\mathbf{F}^{spa}_l$ are processed through $1\times1$ convolution followed by $3\times3$ depth-wise convolution. After that, cross-attention is computed to generate $F_l$, which is the feature enhanced by the low-frequency domain information.

\subsection{Selective feature Fusion Module (SFM)}

$F_h$ is the feature enhanced by the high-frequency domain information, and it primarily captures the texture and edges of the image. $F_l$ is the feature enhanced by the low-frequency domain information, and it primarily represents the contextual and structural information of the image. Feature fusion is essential to obtain a comprehensive representation of the input data. As illustrated in Fig. \ref{fig3}, we employ a spatial selection mechanism to fuse $F_h$ and $F_l$. By dynamically adjusting the attention weights based on the characteristics of the input data, the most relevant feature components are extracted for feature representation.

First, for the two different features $F_h$ and $F_l$, we perform a concatenation operation along the channel dimension. Then, we apply average pooling and max pooling to the concatenated features to capture the spatial relationships of both features, obtaining smooth global information and refined local information as follows:
\begin{equation}
F_c = \text{Concat}(F_h, F_l)
\end{equation}
\begin{equation}
U_{avg} = \text{avgpool}(F_c); \quad U_{max} = \text{maxpool}(F_c)
\end{equation}
Here $F_c$ represents the aggregated features obtained from concatenation, and avgpool and maxpool denote the average pooling and max pooling operations, respectively.

Next, to enable information exchange between the two types of features, we concatenate the two spatially pooled features $U_{avg}$ and $U_{max}$, resulting in a representation with two channels. A $7\times7$ convolutional layer is then applied to further extract joint features, generating an attention map suitable for feature selection. The sigmoid activation function is used to produce the spatially attention mask as follows:
\begin{equation}
[\Gamma_1, \Gamma_2] = \sigma \left( \text{Conv}_{7\times 7} \left( \text{Concat}(U_{avg}, U_{max}) \right) \right).
\end{equation}
After that, $\Gamma_1$ and $\Gamma_2$ are combined with $F_h$ and $F_l$ through element-wise multiplication as follows:
\begin{equation}
\hat{F} = \text{Conv}_{1\times 1}\left( \Gamma_1 \cdot F_h + \Gamma_2 \cdot F_l \right)
\end{equation}

We employ a residual structure with element-wise multiplication to dynamically adjust the weights of different features in the input $\mathbf{X}_{in}$ as follows:
\begin{equation}
\mathbf{X}_{out} = \hat{F} \cdot \mathbf{X}_{in}.
\end{equation}

The SFM effectively fuses features from different sources, while the residual structure introduces non-linear transformation. This enables the final feature map to encompass both rich semantic information of spatial distributions and detailed edge features.

\subsection{Loss Function}
We employ a composite loss function, $\mathcal{L}_{total}$, which combines Cross-Entropy loss $\mathcal{L}_{ce}$ and Dice loss $\mathcal{L}_{dice}$ to balance pixel-wise classification accuracy and regional segmentation performance:

\begin{equation}
\mathcal{L}_{total} = \mathcal{L}_{ce} + \mathcal{L}_{dice}.
\end{equation}

The Cross-Entropy loss is defined as:

\begin{equation}
\mathcal{L}_{ce} = -\frac{1}{N} \sum_{n=1}^{N} \sum_{k=1}^{K} y_k^n \log \hat{y}_k^n.
\end{equation}
$\mathcal{L}_{ce}$ measures the discrepancy between predicted class probabilities $\hat{y}_k^n$ and true labels $y_k^n$ for $N$ samples across $K$ classes.

To address the class imbalance, we incorporate Dice loss:
\begin{equation}
\mathcal{L}_{dice} = -\frac{2}{N} \sum_{n=1}^{N} \sum_{k=1}^{K} \frac{\hat{y}_k^n y_k^n}{\hat{y}_k^n} + y_k^n.
\end{equation}

Dice loss emphasizes high-confidence predictions, thereby enhancing the model's performance under class imbalance. This combined approach improves both pixel classification accuracy and segmentation quality by optimizing the overlap between predicted and true regions.

\section{Experimental Results and Analysis} 

This section details the experimental setup, including datasets, evaluation metrics, and implementation specifics. Subsequently, we present a series of ablation studies, comparative evaluations, and analyses of model complexity to highlight the contributions of key components within AFENet, demonstrating its superior performance.
\subsection{Experimental Settings}
\subsubsection{Datasets}
We conducted experiments on three widely used datasets: ISPRS Vaihingen, ISPRS Potsdam, and LoveDA \cite{LOVEDA}. These datasets serve as benchmarks for semantic segmentation in remote sensing imagery, encompassing diverse land cover types, environmental conditions, and scene complexities. By leveraging these datasets, we ensure the broad applicability and comparability of our method against existing state-of-the-art methods.

\textbf{ISPRS Vaihingen:} The Vaihingen dataset contains 33 orthophotos from the Vaihingen region in Germany, with an average image size of $2494 \times 2064$ pixels and a Ground Sampling Distance (GSD) of 9 cm. Each image includes three spectral bands: near-infrared, red, and green, along with corresponding Digital Surface Models (DSM) and normalized DSM (NDSM). The dataset is annotated into six semantic classes: impervious surfaces, buildings, low vegetation, trees, cars, and clutter/background. To maintain consistency with previous studies, we utilized 16 training images specified by the ISPRS competition, while the remaining 17 images were reserved for testing. Images were preprocessed by cropping them into non-overlapping patches of $1024 \times 1024$ pixels.

\textbf{ISPRS Potsdam:} The Potsdam dataset comprises 38 orthophotos from Potsdam, Germany, each with a size of $6000 \times 6000$ pixels and a GSD of 5 cm. These images feature four spectral bands (red, green, blue, and near-infrared) alongside DSMs and true orthophotos (TOP). Following the ISPRS competition guidelines, we employed a subset of 23 images (IDs: 2\_10, 2\_11, 2\_12, 3\_10, 3\_11, 3\_12, 4\_10, 4\_11, 4\_12, 5\_10, 5\_11, 5\_12, 6\_7, 6\_8, 6\_9, 6\_10, 6\_11, 6\_12, 7\_7, 7\_8, 7\_9, 7\_11, and 7\_12) for training, with the remaining 15 images used for testing. For computational efficiency, all images were divided into patches of $1024 \times 1024$ pixels.

\textbf{LoveDA:} The LoveDA dataset consists of 5,987 high-resolution (0.3GSD) remote sensing images from urban and rural areas in Nanjing, Changzhou, and Wuhan, each measuring $1024\times1024$ pixels. The dataset includes annotations for seven semantic classes: buildings, roads, water, barren land, forest, agriculture, and background. To account for domain-specific challenges such as multi-scale objects, intricate backgrounds, and class imbalance, we adopted the competition-defined splits: 2,522 images for training, 1,669 for validation, and 1,796 for testing. This standardized partitioning ensures consistent evaluation across diverse scenes and scales.

\subsubsection{Evaluation Metrics}
To evaluate the segmentation performance, we employed widely recognized metrics: Overall Accuracy (OA), mean F1 Score(mF1), and Mean Intersection over Union (mIoU). The definitions of these metrics are as follows:
\vspace{-0.1em}
\begin{equation}
\text{OA} = \frac{\sum_{i=1}^{N} \text{TP}_i}{\sum_{i=1}^{N} \left(\text{TP}_i + \text{FP}_i + \text{FN}_i + \text{TN}_i\right)},
\end{equation}

{
\begin{equation}
\text{Precision} = \frac{1}{N} \sum_{i=1}^{N} \frac{\text{TP}_i}{\text{TP}_i + \text{FP}_i},
\end{equation}
\vspace{-0.1em}
\begin{equation}
\text{Recall} = \frac{1}{N} \sum_{i=1}^{N} \frac{\text{TP}_i}{\text{TP}_i + \text{FN}_i},
\end{equation}

\begin{equation}
\text{mF1} = \frac{2 \times (\text{Precision} \times \text{Recall})}{\text{Precision} + \text{Recall}},
\end{equation}
\vspace{-0.1em}
\begin{equation}
\text{mIoU} = \frac{1}{N} \sum_{i=1}^{N} \frac{\text{TP}_i}{\text{TP}_i + \text{FP}_i + \text{FN}_i},
\end{equation}
}

where $\text{TP}_i$, $\text{FP}_i$, $\text{TN}_i$, and $\text{FN}_i$ denote the true positives, false positives, true negatives, and false negatives for class $i$, respectively. $N$ represents the total number of classes (excluding the clutter/background category).

In addition to these average metrics, we also compute the F1 Score and IoU for each individual class. These per-class metrics are presented in the experimental results section to provide a more granular evaluation of the model's performance across different categories.

\subsubsection{Implementation Details}
The experiments were conducted on a system running Ubuntu 20.04 with an NVIDIA GTX 4090 GPU, utilizing the PyTorch 2.0.0 framework. The Adam optimizer was employed with an initial learning rate of 0.0006 and a weight decay of 0.01. 

For both the Vaihingen and Potsdam datasets, images were randomly cropped into $512 \times 512$ patches. Data augmentation techniques, including random scaling, random vertical and horizontal flips, were applied during training. The batch size was set to 8, with each epoch comprising 100 batches. During testing, test-time augmentation such as horizontal and vertical flips was used to enhance robustness.

Additionally, to evaluate model complexity, we calculated floating-point operations (FLOPs) and the number of parameters required to process a $3 \times 512 \times 512$ input image, providing insights into computational efficiency.

\begin{table}[ht]
\centering
\caption{Ablation study on the ISPRS Vaihingen dataset of the proposed AFENet}
\vspace{-1mm}
\label{t_ab_1}
\begin{tabular}{l|cccc}
\toprule\midrule
\multicolumn{1}{c|}{Method} & Params & FLOPs & mIoU (\%) & mF1 (\%) \\ 
\midrule
AFENet & 20.23M & 25.59G & 84.55 & 91.54 \\ 
AFENet w/o AFSIM-Low & 19.70M & 25.27G & 83.49 & 91.03 \\ 
AFENet w/o AFSIM-High & 19.70M & 25.27G & 83.38 & 90.97 \\ 
AFENet w/o AWM+Fixed & 19.86M & 25.36G & 83.63 & 91.09 \\ 
AFENet w/o SFM+Add & 20.20M & 25.52G & 83.97 & 91.21 \\ 
\midrule\bottomrule
\end{tabular}
\vspace{-5mm}
\end{table}

\begin{table}[ht]
\centering
\caption{Ablation study of AFENet with different component combinations}
\vspace{-2mm}
\label{ab_ex_t2}
\resizebox{\columnwidth}{!}{%
\begin{tabular}{cccc|ccc}
\toprule \midrule
Baseline & AFSIM-Low & AFSIM-High & SFM & Params & FLOPs & mIoU (\%) \\
\midrule
\checkmark &  &  &  & 17.82M & 24.16G & 82.62  \\ 
\checkmark & \checkmark &  &  & 19.67M & 25.20G & 83.35  \\ 
\checkmark &  & \checkmark &  & 19.67M & 25.20G & 83.46  \\ 
\checkmark &  &  & \checkmark & 19.10M & 24.93G & 83.20  \\ 
\checkmark & \checkmark & \checkmark &  & 20.20M & 25.52G & 83.92  \\ 
\checkmark & \checkmark & \checkmark & \checkmark & 20.23M & 25.59G & 84.55  \\ 
\midrule \bottomrule
\end{tabular}%
}
\vspace{-5mm}
\end{table}

\begin{figure}[H]
\centering  
\includegraphics[width=0.46\textwidth]{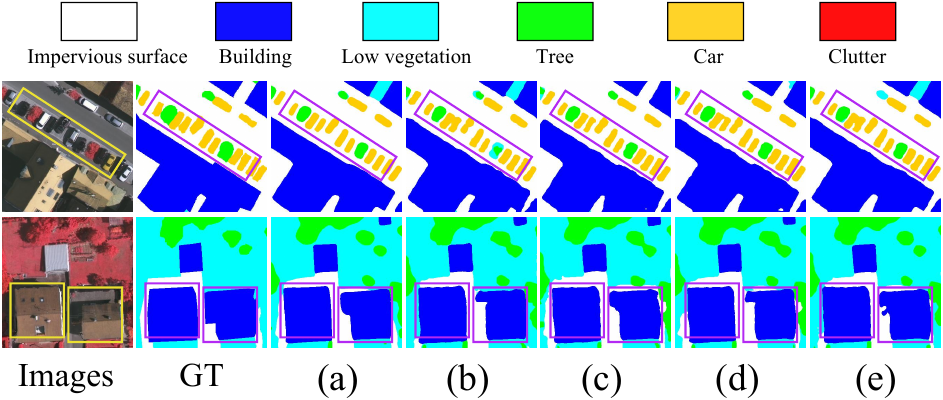} 
\vspace{-1mm}
\caption{Visualization of segmentation performance after removing different components from AFENet on the ISPRS Vaihingen dataset, focusing on magnified local areas. (a) represents AFENet; (b) represents AFENet without AFSIM-Low; (c) represents AFENet without AFSIM-High; (d) represents AFENet with AWM+Fixed; and (e) represents AFENet with SFM+Add.}
\label{ab_ex_fig1}
\vspace{-5mm}
\end{figure}

\vspace{-1mm}

\begin{figure}[H]
\centering  
\includegraphics[width=0.46\textwidth]{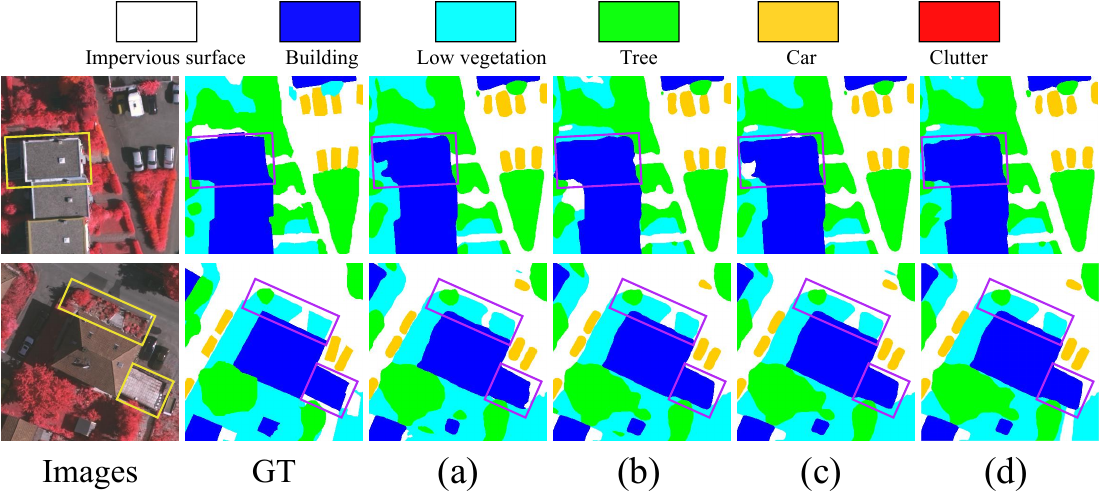}    
\vspace{-2mm}
\caption{Visualization of segmentation performance for different AFSIM configurations on the ISPRS Vaihingen dataset, focusing on magnified local areas. (a) Baseline. (b) Baseline with the low-frequency branch of AFSIM. (c) Baseline with the high-frequency branch of AFSIM. (d) Baseline with the complete AFSIM.}
\label{ab_ex_fig2}
\vspace{-5mm}
\end{figure}
\vspace{-1mm}
\begin{figure}[H]
\centering  
\includegraphics[width=0.46\textwidth]{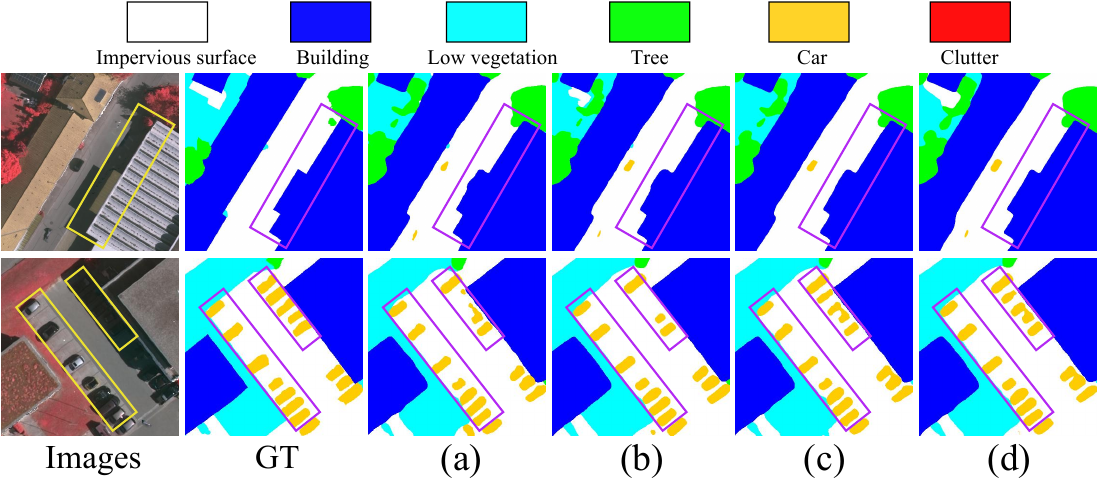}
\vspace{-2mm}
\caption{Visualization of segmentation performance for AFENet with different combinations incorporating SFM on the ISPRS Vaihingen dataset, focusing on magnified local areas. (a) Baseline. (b) Baseline with SFM. (c) Baseline with AFSIM, and (d) Proposed AFENet.}
\vspace{-5mm}
\label{ab_ex_fig3}
\end{figure}

\subsection{Ablation Experiments}
\label{sec:ablation}
To assess the contributions of individual components in the proposed AFENet, we conducted a series of ablation experiments on the ISPRS Vaihingen dataset. The evaluation primarily focuses on two key performance metrics: mIoU and mean F1 score. Components removed during ablation are denoted by “(w/o)”, while added components are indicated with “(+)”. All results are averaged over multiple runs to ensure robustness and reliability.

\subsubsection{Components of AFENet}

The role of each module in the AFEB was analyzed through ablation experiments. Table \ref{t_ab_1} reports the performance of AFENet after removing specific components, along with the corresponding model parameters and computational complexity. Specifically, “AFSIM-Low” refers to the exclusion of the low-frequency branch in the AFSIM, while “AFSIM-High” indicates the removal of the high-frequency branch. These changes eliminate the interaction and fusion of frequency-domain features with corresponding spatial-domain features. “AWM+Fixed” involves replacing the AWM with a fixed-size window mask for frequency separation, while “SFM+Add” removes the SFM, replacing it with a simple addition operation between feature sets. The results show that removing any module leads to a decline in model performance.
\begin{figure*}[ht!]
\centering
\vspace{-5mm}
\includegraphics [width=5.5in]{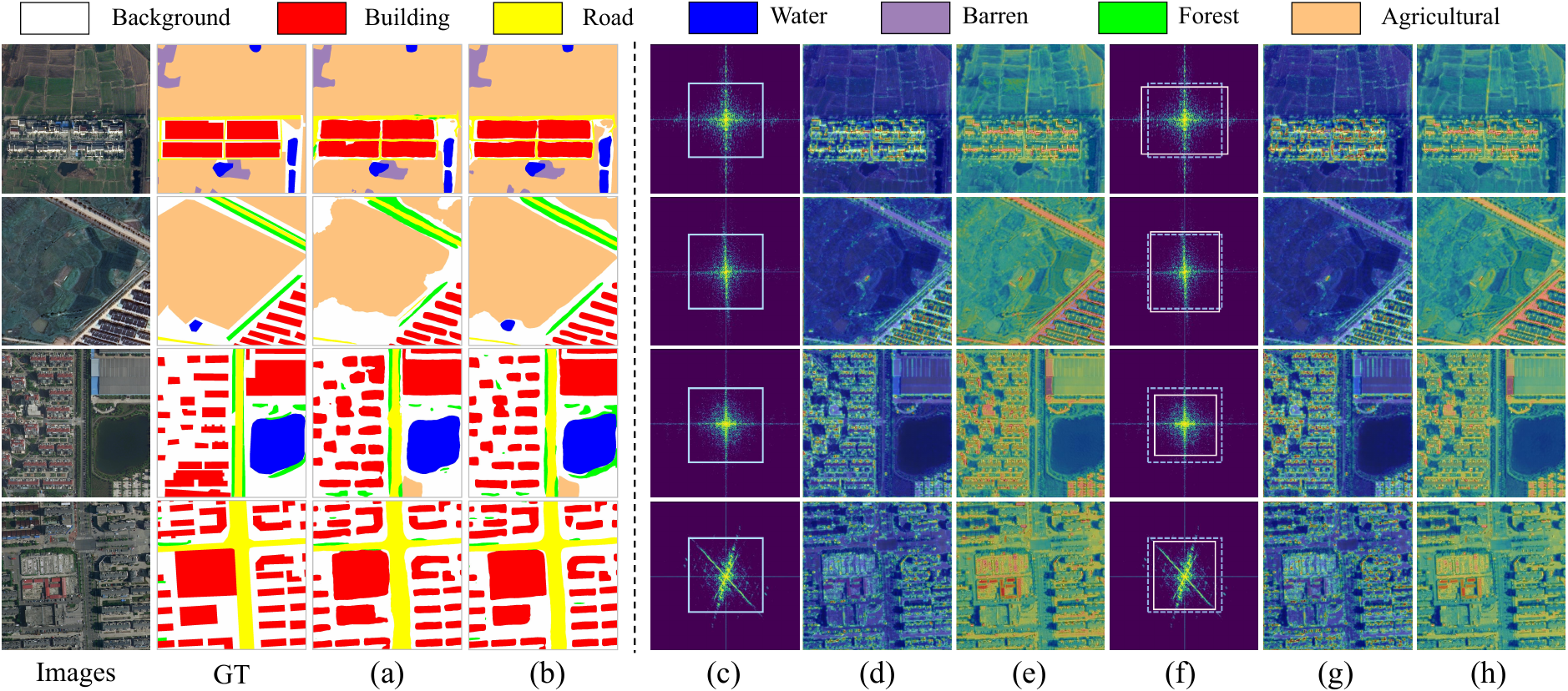}
\caption{
Visual comparison of segmentation performance (left) and window mechanism on the frequency-domain features (right). (a) and (b) present the segmentation results using ``Fixed window" and ``Adaptive window," respectively. (c) shows the FFT spectrum with the blue box indicating the fixed window range. (d) and (e) depict high- and low-frequency features extracted using the fixed window. In (f), the blue dashed line represents the fixed window range, while the white box shows the adaptive window range by AWM. (g) and (h) illustrate high- and low-frequency features extracted using the adaptive window.}
\vspace{-3mm}
\label{ab_ex_fig4}
\vspace{-1mm}
\end{figure*}

Fig. \ref{ab_ex_fig1} visualizes the segmentation results after removing individual modules. The exclusion of the high-frequency branch from AFSIM significantly reduces the model's ability to segment small objects, such as cars, while removing the low-frequency branch compromises the extraction of semantic information from large objects like buildings. Using a fixed-size window instead of AWM diminishes the model's ability to handle objects of varying scales, and the absence of SFM results in suboptimal feature fusion, further degrading performance.

To validate the individual and combined contributions of each component, we progressively added key modules to a baseline model constructed by removing all other components. This baseline retained only the ResNet-18 encoder and the Transformer Block-based decoder, relying exclusively on spatial-domain features. The performance results are summarized in Table \ref{ab_ex_t2}.

Fig. \ref{ab_ex_fig2} highlights the impact of adding different branches of AFSIM (AFSIM-Low and AFSIM-High) to the baseline model, demonstrating the effectiveness of frequency-domain methods in extracting contextual features and enhancing edge details. Fig. \ref{ab_ex_fig3} illustrates the role of SFM in improving global contextual coherence and local detail precision. These results confirm that each proposed component significantly enhances performance, with their integration leading to further improvements.

\begin{table}[h]
\centering
\vspace{-2mm}
\caption{Performance Comparison of Fixed and Adaptive Window}
\vspace{-2mm}
\label{ab_ex_t3}
\resizebox{0.9\columnwidth}{!}{%
\begin{tabular}{c|cccc}
\toprule\midrule
Method       & Params & FLOPs &mIoU (\%) & mF1 (\%) \\ \midrule
Fixed Window  & 20.18M     & 25.58G       &73.80       &84.77     \\
Adaptive Window       & 20.23M      & 25.59G              & 75.18       & 85.68    \\ 
\midrule\bottomrule
\end{tabular}}
\end{table}
\vspace{-2mm}
\begin{table}[h]
\centering

\caption{Ablation Study on Dimensional Compression Methods}
\label{ab_ex_t4}
\resizebox{0.95\columnwidth}{!}{%
\begin{tabular}{c|ccc|ccc}
\toprule
\toprule
\multicolumn{1}{c}{\multirow{2}{*}{\centering Method}} &
\multicolumn{3}{c}{Intermediate Dim $d$} & \multicolumn{3}{c}{Performance} \\
\cmidrule(lr){2-4} \cmidrule(lr){5-7}
 & 512→ & 256→ & 128→ & Params(M) & mIoU(\%) & mF1(\%) \\ 
\midrule
Square Root & 23 & 16 & 11 & \textbf{20.204} & 84.48 & 91.48 \\ 
Fixed//16 & 32 & 16 & 8 & 20.208 & 84.49 & 91.50 \\ 
Fixed//8 & 64 & 32 & 16 & 20.239 & 84.52 & 91.52 \\ 
\textbf{Log-Scale} & 56 & 32 & 18 & 20.230 & \textbf{84.55} & \textbf{91.54} \\ 
\bottomrule
\bottomrule
\end{tabular}
}
\vspace{-2mm}
\end{table}
\subsubsection{Impacts of AFSIM}

To evaluate the effectiveness of AFSIM in separating and utilizing high- and low-frequency features, experiments were conducted by incrementally incorporating its branches (AFSIM-Low and AFSIM-High) and the complete module into the baseline model. As shown in Fig. \ref{ab_ex_fig2}, each component contributes distinct advantages. In Fig. \ref{ab_ex_fig2}(b), incorporating the low-frequency branch improves the model’s ability to extract contextual information, resulting in better recognition of large-scale structures like buildings. However, edge detail smoothness remains limited. In Fig. \ref{ab_ex_fig2}(c), adding the high-frequency branch enhances the segmentation precision of small objects and vegetation edges by filtering shadow effects, though overall contextual extraction remains constrained. Finally, in Fig. \ref{ab_ex_fig2}(d), the integration of both branches leverages their complementary strengths, achieving superior segmentation performance. As detailed in Table \ref{ab_ex_t2}, the complete AFSIM improves mIoU by 0.62\% and 0.51\% compared to using AFSIM-Low or AFSIM-High alone, respectively.

\subsubsection{Impacts of SFM}
We designed SFM to facilitate effective fusion of context and detail features, enhancing the model's expressive capacity. Table \ref{ab_ex_t2} shows that adding SFM to the baseline model increases mIoU by 0.58\%, and further adding SFM to the model containing AFSIM increases mIoU by 0.63\%. Fig. \ref{ab_ex_fig3} illustrates the benefits of SFM in different scenarios: (a) and (b) show improvements in building edge segmentation when SFM is integrated into the baseline model. (c) and (d) highlight enhanced detail accuracy and segmentation consistency when SFM is added to the model containing AFSIM. These results underscore the advantages of SFM in feature fusion and its effectiveness in handling complex scenarios, particularly for fine-grained segmentation tasks.

\begin{table*}[ht]
\centering
\vspace{-5mm}
\caption{Experimental results of different methods on the ISPRS Vaihingen Dataset}
\vspace{-2mm}
\label{vai_ex_t}
\resizebox{0.75\textwidth}{!}{%
\begin{tabular}{@{}c|ccccc|ccc@{}}
\toprule \midrule
Method & Imp. surf & Building & Low veg. & \hspace{2mm}Tree\hspace{2mm} & \hspace{2mm}Car\hspace{2mm} & \hspace{1mm}mF1 (\%) & mIoU (\%) & OA (\%) \\ 
\midrule
FCN8s \cite{long15fcn} & 95.48 & 92.12 & 82.03 & 88.80 & 79.96 & 87.68 & 78.55 & 89.22 \\ 
UNet \cite{ronneberger2015u} & 96.29 & 94.75 & 82.87 & 88.99 & 84.04 & 89.39 & 81.25 & 90.21 \\ 
DeepLabv3+ \cite{chen2018encoder} & 96.64 & 95.40 & 83.16 & 89.14 & 85.15 & 89.90 & 82.08 & 90.78 \\ 
BANet \cite{wang2021transformer} & 96.58 & 95.32 & 83.27 & 89.54 & 88.29 & 90.60 & 83.18 & 90.83 \\ 
MA-Net \cite{9487010} & 96.63 & 95.43 & 84.15 & 89.79 & 86.61 & 90.52 & 83.05 & 90.98 \\ 
A$^2$-FPN \cite{li2022a2} & 96.59 & 95.30 & 83.27 & 89.37 & 86.93 & 90.29 & 82.68 & 90.78 \\ 
UNetFormer \cite{wang2022unetformer} & 96.80 & 95.49 & 83.61 & 89.36 & 87.62 & 90.57 & 83.14 & 90.99 \\ 
CMTFNet \cite{10247595} & 96.74 & 95.70 & 84.23 & 89.76 & 88.31 & 90.95 & 83.74 & 91.15 \\ 
SFFNet \cite{Yang2024SFFNet} & 96.86 & 95.57 & 84.79 & 90.13 & 89.27 & 91.32 & 84.38 & 91.36 \\ 
SparseFormer \cite{10817638} & 96.72 & 95.65 & 84.42 & 89.93 & 88.85 & 91.11 & 84.12 & 91.21 \\
XNet \cite{10376766} & 96.81 & 95.63 & 84.96 & 90.32 & 89.29 & 91.40 & 84.43 & 91.52 \\

\midrule
Proposed AFENet & \textbf{96.90} & \textbf{95.72} & \textbf{85.07} & \textbf{90.64} & \textbf{89.37} & \textbf{91.54} & \textbf{84.55} & \textbf{91.67} \\ 
\midrule \bottomrule
\end{tabular}%
}
\vspace{-2mm}
\end{table*}

\begin{table*}[ht]
\centering
\vspace{0mm}
\caption{Experimental results of different methods on the ISPRS Potsdam Dataset}
\vspace{-2mm}
\resizebox{0.75\textwidth}{!}{%
\begin{tabular}{@{}c|ccccc|ccc@{}}
\toprule\midrule
Method & Imp. surf & Building & Low veg. & \hspace{2mm}Tree\hspace{2mm} & \hspace{2mm}Car\hspace{2mm} & \hspace{1mm}mF1 (\%) & mIoU (\%) & OA (\%) \\ 
\midrule
FCN8s\cite{long15fcn} & 90.95 & 92.84 & 85.33 & 86.25 & 92.80 & 89.64 & 81.37 & 88.34 \\ 
UNet\cite{ronneberger2015u} & 92.66 & 94.89 & 86.17 & 87.71 & 95.36 & 91.35 & 84.36 & 89.94 \\ 
DeepLabv3+\cite{chen2018encoder} & 93.15 & 95.43 & 86.45 & 87.78 & 95.21 & 91.40 & 84.42 & 90.07 \\ 
BANet\cite{wang2021transformer} & 93.54 & 96.23 & 87.06 & 88.45 & 95.45 & 92.15 & 85.62 & 90.89 \\ 
MA-Net\cite{9487010} & 93.24 & 95.64 & 86.83 & 88.15 & 95.83 & 91.94 & 85.31 & 90.82 \\ 
A$^2$-FPN\cite{li2022a2} & 93.43 & 96.12 & 86.89 & 88.38 & 95.17 & 92.00 & 85.53 & 90.83 \\ 
UNetFormer\cite{wang2022unetformer} & 93.78 & 96.20 & 87.52 & 88.44 & 96.17 & 92.42 & 86.14 & 91.21 \\ 
CMTFNet\cite{10247595} & 94.01 & 96.46 & 87.32 & 88.39 & 96.21 & 92.48 & 86.58 & 91.37 \\ 
SFFNet\cite{Yang2024SFFNet} & 94.20 & 96.62 & 87.73 & 89.20 & 96.33 & 92.82 & 86.93 & 91.72 \\ 
SparseFormer \cite{10817638} & 93.96 & 96.50 & 87.55 & 89.20 & 96.15 & 92.67 & 86.83 & 91.56 \\
XNet \cite{10376766} & 94.25 & 96.57 & 87.83 & 89.43 & 96.43 & 92.90 & 87.14 & 91.83 \\
\midrule
Proposed AFENet & \textbf{94.57} & \textbf{96.86} & \textbf{88.17} & \textbf{89.78} & \textbf{96.83} & \textbf{93.24} & \textbf{87.50} & \textbf{92.03} \\ 
\midrule \bottomrule
\end{tabular}%
}
\label{pot_ex_t}
\vspace{-2mm}
\end{table*}

\subsubsection{Impacts of AWM}

In AFSIM, we designed an Adaptive Window-mask Module (AWM) to adaptively separate high- and low-frequency information in the frequency domain, enabling the model to effectively process diverse remote sensing images. The dual fully-connected layers ($\text {FC}_1$ and $\text {FC}_2$) in AWM form an adaptive threshold learning pathway: $\text {FC}_1$ filters channel features by compressing from $C$ to dynamic intermediate dimension $d$, preserving texture patterns while suppressing noise, and $\text {FC}_2$ subsequently projects these refined features to spectral thresholds $[r_h, r_w]$ that control frequency boundaries along spatial axes.

To validate the dimension compression strategy, we conducted ablation studies comparing four approaches: Square Root, Fixed//8, Fixed//16, and our Log-Scale method. As shown in Table~\ref{ab_ex_t4}, the Log-Scale compression demonstrates superior gradient stability compared to Square Root while maintaining 0.4\% fewer parameters than Fixed//8 on the Vaihingen dataset. This more reasonable dimension compression reduces redundant parameters and enhances performance.

To further evaluate the efficacy of AWM in frequency separation, experiments were conducted on the LoveDA dataset. Table \ref{ab_ex_t3} presents the performance comparison of two methods on the LoveDA validation set: “Fixed Window” employs a fixed-size mask window with dimensions $\frac{H}{2} \times \frac{W}{2}$, while “Adaptive Window” utilizes a dynamically sized window generated by AWM. Fig. \ref{ab_ex_fig4} provides a visual comparison of segmentation performance (left) and illustrates the impact of the two window mechanisms on the frequency spectrum and corresponding frequency-domain features (right).

In the first and second rows, which depict rural regions characterized by large homogeneous areas such as farmland and wasteland, the adaptive window dynamically adjusts to prioritize essential low-frequency information while effectively incorporating high-frequency details. This adjustment enhances the segmentation accuracy of farmland and building edges compared to the fixed window approach.

Conversely, the third and fourth rows represent urban regions with more complex land-use patterns, including intricate structures such as buildings and road edges. In these scenarios, segmentation relies more heavily on high-frequency information. The adaptive window captures additional high-frequency details by reducing the retention of redundant low-frequency components, whereas the fixed window tends to overemphasize low-frequency features, thereby limiting the network’s ability to fully exploit high-frequency information for detailed segmentation tasks.

Overall, AFENet, equipped with the AWM, demonstrates remarkable adaptability to varying land-use scenarios, particularly in heterogeneous environments. By dynamically balancing the extraction of high- and low-frequency features, the proposed module enhances the utilization of frequency-domain information, leading to superior segmentation performance across diverse and complex remote sensing datasets.

\begin{figure*}[ht]
\centering
\vspace{-5mm}
\includegraphics [width=5in]{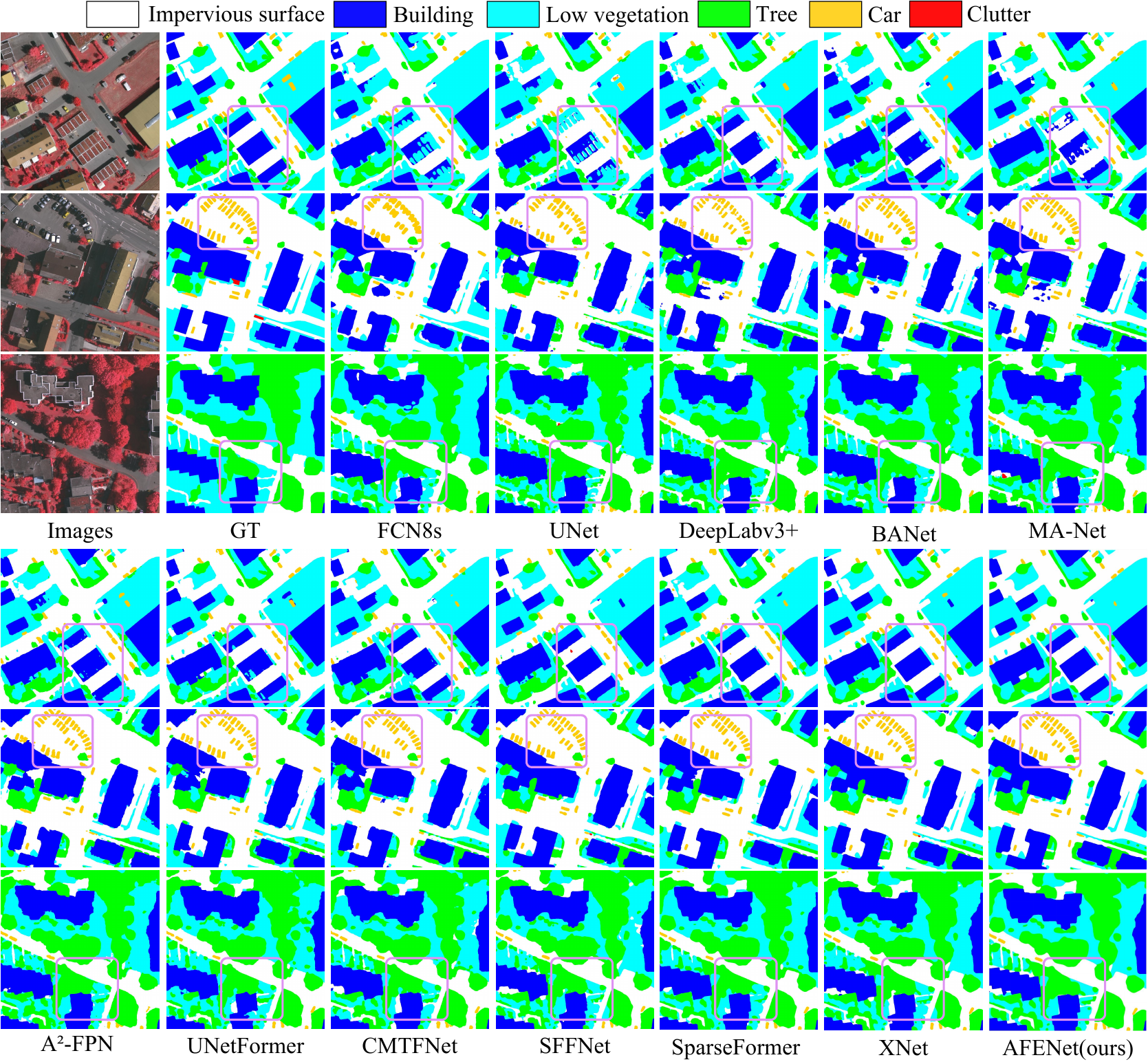}
\vspace{-2mm}
\caption{Experimental result visualization of different methods on the ISPRS Vaihingen dataset.}
\vspace{-4mm}
\label{vai_ex}
\end{figure*}

\subsection{Comparative Experiments}

We compared our proposed method, AFENet, against several state-of-the-art (SOTA) methods, including FCN8s \cite{long15fcn}, UNet \cite{ronneberger2015u}, DeepLabv3+ \cite{chen2018encoder}, BANet \cite{wang2021transformer}, MA-Net \cite{9487010}, A
2
-FPN \cite{li2022a2}, UNetFormer \cite{wang2022unetformer}, CMTFNet \cite{10247595}, SFFNet \cite{Yang2024SFFNet}, SparseFormer \cite{10817638}, and XNet \cite{10376766}. Among these methods, the first three are predominantly designed for semantic segmentation of natural images, while the latter nine are tailored for remote sensing image segmentation and are recognized as SOTA models. Notably, UNetFormer balances global and local spatial feature extraction through multi-head position attention, CMTFNet employs a Transformer decoder with cross-modality interaction to enhance contextual information extraction, and SFFNet leverages frequency-domain features through wavelet transforms. Specifically, SparseFormer pioneers dual-CNN expert guidance to synergize CNN's local detail capture with Transformer's long-range dependency modeling, while XNet establishes multi-frequency analysis framework through wavelet-based low/high-frequency fusion for enhanced image consistency. Compared to these advanced techniques, AFENet consistently outperforms in segmentation accuracy across the ISPRS Vaihingen, ISPRS Potsdam, and LoveDA datasets, demonstrating its adaptability and superior performance in diverse land cover scenarios.

\subsubsection{Results on the Vaihingen Dataset}

Table \ref{vai_ex_t} presents the quantitative evaluation results of various methods on the Vaihingen dataset, demonstrating that AFENet outperforms all compared approaches across all metrics (mF1, mIoU, and OA). To further illustrate the performance differences, Fig. \ref{vai_ex} presents visual comparisons of segmentation results across three distinct scenarios. These scenarios focus on key areas of interest, such as building clusters, small objects, low vegetation, and tree regions, highlighted in purple for clarity.

The results reveal that existing segmentation frameworks like FCN and UNet exhibit general adaptability across different remote sensing tasks. However, their simpler network structures constrain their ability to effectively capture the rich semantics of remote sensing images. DeepLabv3+ excels in capturing fine details due to its atrous separable convolution-based ASPP and decoder modules. Models such as A$^2$-FPN effectively incorporate multi-scale context using feature pyramid networks augmented by attention mechanisms, while BANet and MA-Net leverage diverse attention strategies to enhance both global and local information processing. UNetFormer strikes a balance between convolutional feature extraction and contextual attention, achieving consistent performance across diverse categories. CMTFNet advances feature interaction between channels, improving contextual representation, while SFFNet refines segmentation boundaries by emphasizing frequency-domain features. Notably, SparseFormer achieves multi-category superiority through its dual-branch architecture that synergizes CNN-based local pattern capture with Transformer's global dependency modeling, whereas XNet establishes cross-frequency understanding via wavelet-based feature decomposition and fusion.

In contrast, AFENet stands out in scenarios demanding precise edge delineation, such as identifying small objects like vehicles or detailed building structures. This success is attributed to its dual approach of separating and selectively fusing high-frequency and low-frequency information. Compared to other frequency-domain methods like SFFNet and XNet, AFENet better accounts for the interaction and fusion of spatial-domain features, providing more accurate vegetation and tree boundaries, generating segmentation outputs closely aligned with ground truth annotations.

\begin{figure*}[ht]
\centering
\vspace{-5mm}
\includegraphics [width=5in]{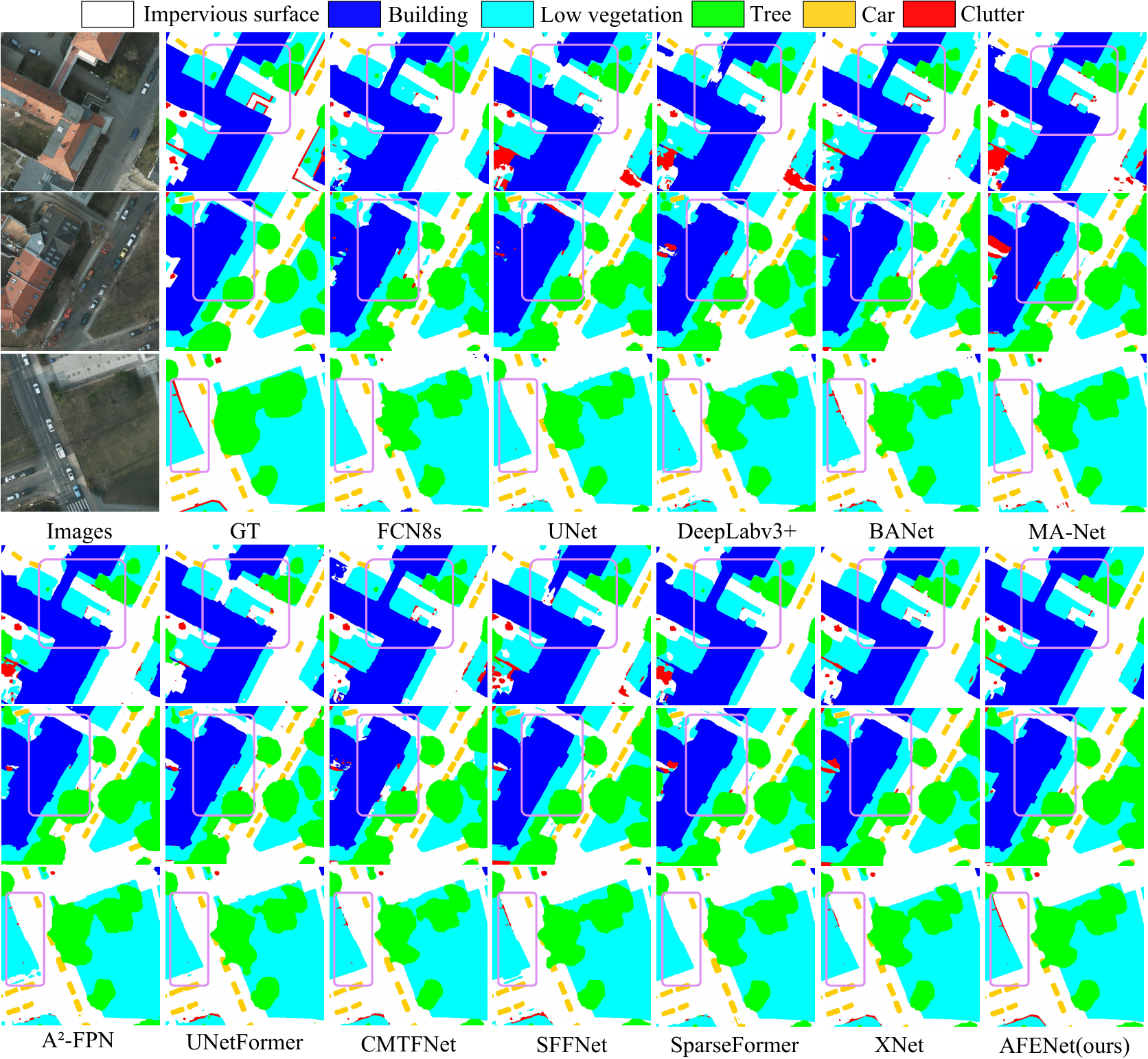}
\vspace{-2mm}
\caption{Experimental result visualization of different methods on the ISPRS Potsdam Dataset.}
\vspace{-4mm}
\label{pot_ex}
\end{figure*}

\begin{table*}[h!]
\centering
\caption{Experimental results of different methods on the LoveDA dataset}
\vspace{-1mm}
\resizebox{0.75\textwidth}{!}{%
\begin{tabular}{@{}c|ccccccc|c@{}}
\toprule\midrule
Method & Background & Building & \hspace{1mm}Road\hspace{1mm} & \hspace{1mm}Water\hspace{1mm} & \hspace{1mm}Barren\hspace{1mm} & \hspace{1mm}Forest\hspace{1mm} & Agricultural & \hspace{1mm}mIoU (\%)\hspace{1mm} \\ 
\midrule
FCN8s\cite{long15fcn} & 42.91 & 50.33 & 46.82 & 76.38 & 17.76 & 44.39 & 57.71 & 48.04 \\ 
UNet\cite{ronneberger2015u} & 43.38 & 53.53 & 53.52 & 76.03 & 17.75 & 44.74 & 58.33 & 49.61 \\ 
DeepLabv3+\cite{chen2018encoder} & 45.06 & 56.28 & 56.37 & 79.31 & 16.43 & 43.85 & 62.32 & 51.38 \\ 
BANet\cite{wang2021transformer} & 45.35 & 58.13 & 51.84 & 79.79 & 18.18 & 46.35 & 61.25 & 51.56 \\ 
MA-Net\cite{9487010} & 44.33 & 56.17 & 53.28 & 79.24 & 20.51 & 45.49 & 60.36 & 51.34 \\ 
A$^2$-FPN\cite{li2022a2} & 44.84 & 57.25 & 55.68 & 78.73 & 20.79 & 46.10 & 62.68 & 52.29 \\ 
UNetFormer\cite{wang2022unetformer} & 44.59 & 58.82 & 55.36 & 79.92 & 19.10 & 46.15 & 61.31 & 52.18 \\ 
CMTFNet\cite{10247595} & 46.20 & 58.57 & 56.06 & 79.60 & 19.88 & 46.60 & 63.83 & 52.96 \\ 
SFFNet\cite{Yang2024SFFNet} & 47.40 & 59.13 & 58.16 & 81.43 & 21.05 & 47.74 & 65.32 & 54.32 \\ 
SparseFormer \cite{10817638} & 47.12 & 58.77 & 58.03 & 80.78 & 20.42 & 48.32 & 64.74 & 54.03 \\
XNet \cite{10376766} & 46.35 & 59.05 & 58.96 & 81.37 & 21.28 & 48.19 & 65.89 & 54.44 \\
\midrule
\textbf{Proposed AFENet} & \textbf{47.43} & \textbf{59.22} & \textbf{59.17} & \textbf{81.55} & \textbf{21.36} & \textbf{48.68} & \textbf{66.40} & \textbf{54.83} \\ 
\midrule\bottomrule
\end{tabular}%
}
\label{loveda_ex_t}
\end{table*}

\subsubsection{Results on the Potsdam Dataset}
To assess the network's performance across varying scenes and spatial resolutions, additional experiments were conducted on the Potsdam dataset. Table \ref{pot_ex_t} summarizes the quantitative results, where AFENet achieves the highest scores across all metrics, reaffirming its superior performance. Notably, all methods exhibit better performance on the Potsdam dataset compared to the Vaihingen dataset, primarily due to the Potsdam dataset's abundant training data, which provides a more comprehensive representation of land cover variations.

For a deeper understanding of the results, qualitative analyses are presented in Fig. \ref{pot_ex}, with regions of interest, such as complex building clusters and low vegetation areas, highlighted in purple. Despite the advantage of extensive training data in the Potsdam dataset, FCN8s demonstrates the weakest performance, highlighting their limitations in high-resolution imagery. Models like UNet and DeepLabv3+ show limited effectiveness in handling intricate structures, especially in areas with dense and complex buildings. BANet and MA-Net, while better at segmenting large-scale features, fail to capture certain building segments, reducing their overall accuracy. A$^2$-FPN delivers more balanced segmentation results but lacks the fine detail resolution achieved by higher-performing methods.
\begin{figure*}[t]
\centering
\includegraphics [width=5in]{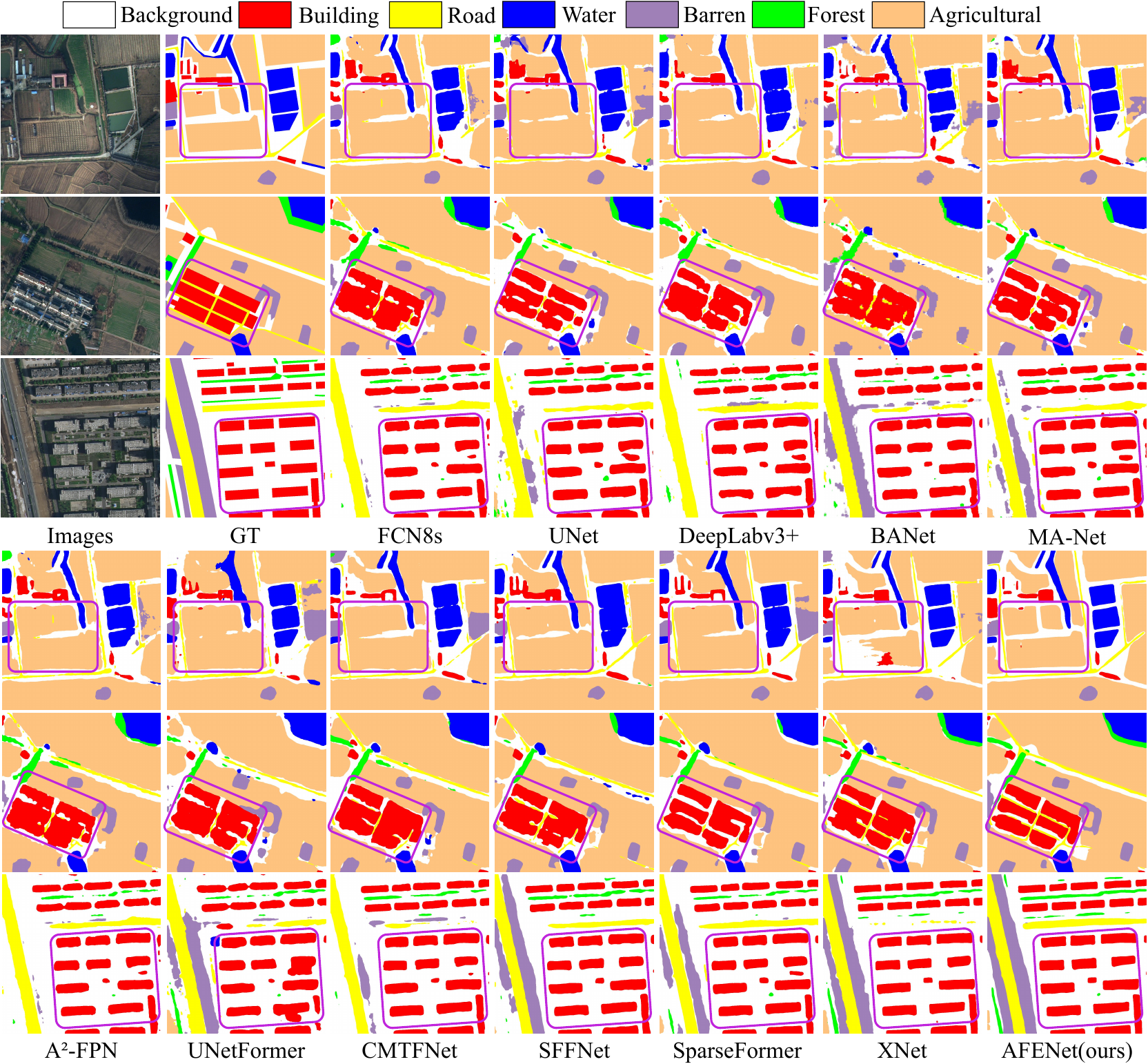}
\vspace{-2mm}
\caption{Experimental result visualization of different methods on the LoveDA Dataset.}
\vspace{-4mm}
\label{loveda_ex}
\end{figure*}
The high spatial resolution of the Potsdam dataset underscores the importance of capturing fine details and maintaining edge precision. Models with a primary focus on spatial features, such as UNetFormer and CMTFNet, struggle to ensure semantic consistency, often resulting in compromised boundary accuracy. Conversely, SFFNet leverages frequency-domain information to enhance detail segmentation, but it suffers from semantic loss in certain areas, particularly low vegetation regions. SparseFormer, although it effectively maintains contextual consistency, exhibits slightly inferior performance in handling local details such as building edges. XNet, which relies solely on frequency features without processing spatial features, is impaired in its ability to semantically identify large-scale objects. In contrast, AFENet achieves smoother edge segmentation for both buildings and low vegetation, outperforming spatially focused models and frequency-centric approaches. By effectively integrating spatial and frequency-domain features, AFENet demonstrates a balanced approach that ensures superior segmentation accuracy and robust performance across diverse scenarios.

\subsubsection{Results on the LoveDA Dataset}
To assess the performance of our method across varying spatial resolutions and diverse land cover scenarios, we conducted comparative experiments on the LoveDA dataset. Table \ref{loveda_ex_t} presents the results on the official test set, presenting the IoU for seven semantic classes and the mIoU. Our proposed AFENet achieves the highest mIoU of 54.83\%, outperforming all other models.

Fig. \ref{loveda_ex} visualizes segmentation results from two rural scenes and one urban scene in the validation set. These scenes focus on key regions of interest, such as agricultural land and roads in rural areas, and complex building clusters in urban areas, which are highlighted in purple for clarity. The extensive land cover in LoveDA provides a rich and diverse training set, enabling conventional models like FCN8s and UNet to perform well in segmenting large, uniform classes such as agriculture and water. Attention-based models, including BANet and MA-Net, leverage contextual information to achieve better segmentation in extensive regions and building areas. Similarly, multi-scale approaches such as A$^2$-FPN and CMTFNet effectively extract features from objects of varying sizes, particularly excelling in complex building layouts. UNetFormer and SparseFormer effectively handle local details and long-range dependencies by leveraging CNNs and attention mechanisms for spatial-domain features. SFFNet and XNet, based on frequency methods, effectively segment agricultural fields and building edges. However, these methods focus on specific aspects of the two types of scenes and cannot simultaneously meet the segmentation requirements of multiple scenes.

Our AFENet, however, demonstrates a distinct advantage in edge segmentation, particularly for buildings, roads, and agricultural regions. By employing an adaptive frequency separating mechanism, AFENet supplements the segmentation network with corresponding high-frequency and low-frequency information for different scenes, enhancing the model's understanding of the distribution characteristics of ground objects in various scenes. This enables AFENet to more comprehensively capture multi-scale information and simultaneously enhance its ability to recognize fine details. This results in improved semantic coherence and significantly higher edge segmentation accuracy, especially for objects with varied scales and intricate boundaries.

\begin{table}[h!]
\centering
\vspace{-4mm}
\caption{Comparison of the number of parameters and computational complexity with different methods}
\begin{tabular}{c|cccc}
\toprule\midrule
Method & Backbone & Params  & FLOPs & mIoU (\%) \\
\midrule
FCN8s\cite{long15fcn} & VGG-16 & 30.03M & 320.87G & 78.55 \\
UNet\cite{ronneberger2015u} & ResNet-18 & 14.33M & 21.90G & 81.25 \\
DeepLabv3+\cite{chen2018encoder} & ResNet-18 & 12.33M & 18.42G & 82.08 \\
BANet\cite{wang2021transformer}  & ResNet-18 & 12.73M & 13.06G & 83.18 \\
MA-Net\cite{9487010} & ResNet-18 & 11.99M & 22.12G & 83.05 \\
A$^2$-FPN\cite{li2022a2} & ResNet-18 & 22.82M & 41.83G & 82.68 \\
UNetFormer\cite{wang2022unetformer} & ResNet-18 & 11.68M & 11.74G & 83.14 \\
CMTFNet\cite{10247595} & ResNet-50 & 30.07M & 33.07G & 83.74 \\
SFFNet\cite{Yang2024SFFNet} & ConvNext-tiny & 34.16M & 51.67G & 84.38 \\
SparseFormer\cite{10817638} & Swin-tiny & 42.52M & 41.95G & 84.12 \\
XNet\cite{10376766} & Custom & 41.42M & 115.09G & 84.43 \\
\midrule
Proposed AFENet & ResNet-18 & 20.23M & 25.59G & 84.55 \\
\midrule\bottomrule
\end{tabular}
\label{coml_t}
\end{table}

\subsection{Complexity Analysis}

Table \ref{coml_t} presents the comparison of computational complexity, parameter count, and mIoU performance on the Vaihingen dataset across different methods, all evaluated under consistent settings. To ensure fairness, we used pre-trained VGG-16 and ResNet-18 as backbones to compare classic semantic segmentation models, such as FCN8s, UNet, and Deeplabv3+. For comparison with recent SOTA methods, we selected models with similar parameter counts or those also using ResNet-18 as the backbone, such as BANet, A$^2$-FPN, and UNetFormer. The results in Table \ref{coml_t} show that AFENet effectively balances computational efficiency and segmentation performance. Notably, AFENet achieves superior segmentation performance with lower computational complexity compared to networks with larger backbones, such as ResNet-50 and ConvNext, and more complex parameterizations like XNet.  In the task of remote sensing image semantic segmentation, AFENet demonstrates a better balance between performance and parameter count.

\section{Conclusion} 

In this paper, we propose AFENet for remote sensing image semantic segmentation. We find existing remote sensing image segmentation methods are weak in adapting network parameters to the input data with different land covers. Hence, AFSIM is proposed to tackle this issue. It adaptively modulates the high- and low-frequency components according to the content of the input image. It adaptively generates two masks to separate high- and low-frequency components for each input image, providing optimal details and contextual supplementary information for ground object feature representation. In addition, we design SFM to selectively fuse global context and local detailed features to enhance the network's representation capability. Hence, the interactions between frequency and spatial features are further enhanced. Extensive experiments are implemented on ISPRS Vaihingen, ISPRS Potsdam, and LoveDA datasets, demonstrating that the proposed AFENet significantly outperforms state-of-the-art methods in segmentation accuracy and generalizability. Notably, AFSIM has proven critical in dynamically modulating feature representations to diverse scenarios, ensuring robust performance across rural and urban landscapes with different textures and structural complexities. In the future, we will work on extending AFENet to handle multimodal remote sensing datasets, such as optical and SAR images, to explore its potential in multisource data fusion. Additionally, we plan to optimize the network architecture to enhance segmentation performance while reducing computational costs, enabling its application in unmanned aerial vehicles.

\bibliographystyle{IEEEtran}
\bibliography{source}

\end{document}